\documentclass[reprint,amsmath,amssymb,aps,floatfix]{revtex4-1}
\usepackage{graphicx}
\usepackage{dcolumn}
\usepackage{bm}
\usepackage[hidelinks]{hyperref}
\usepackage[capitalise]{cleveref}
\usepackage{siunitx}
\usepackage{tabularx}
\usepackage[normalem]{ulem}
\usepackage{color}
\usepackage{float}

\usepackage{titlesec}

\titlespacing\section{0pt}{12pt plus 4pt minus 2pt}{0pt plus 2pt minus 2pt}
\titlespacing\subsection{0pt}{12pt plus 4pt minus 2pt}{0pt plus 2pt minus 2pt}
\titlespacing\subsubsection{0pt}{12pt plus 4pt minus 2pt}{0pt plus 2pt minus 2pt}

\begin{document}

\title{Decoherence benchmarking of superconducting qubits}

\author{Jonathan J. Burnett\normalfont\textsuperscript{$\dagger$}
}
\thanks{Present address: National Physical Laboratory, Hampton road, Teddington, UK, TW11 0LW}
\author{Andreas Bengtsson}
\thanks{These two authors contributed equally}
\author{Marco Scigliuzzo}
\author{David Niepce}
\author{Marina Kudra}
\author{Per Delsing}
\author{Jonas Bylander}
\email{bylander@chalmers.se}

\affiliation{Microtechnology and Nanoscience, Chalmers University of Technology, SE-412 96, G\"oteborg, Sweden}

\date{\today}

\begin{abstract}

We benchmark the decoherence of superconducting transmon qubits to examine the temporal stability of energy relaxation, dephasing, and qubit transition frequency. By collecting statistics during measurements spanning multiple days, we find the mean parameters $\overline{T_{1}}$ = \SI{49}{\micro\second} and $\overline{T_{2}^{*}}$ = \SI{95}{\micro\second}; however, both of these quantities fluctuate, explaining the need for frequent re-calibration in qubit setups. Our main finding is that fluctuations in qubit relaxation are local to the qubit and are caused by instabilities of near-resonant two-level-systems (TLS). Through statistical analysis, we determine sub-millihertz switching rates of these TLS and observe the coherent coupling between an individual TLS and a transmon qubit. Finally, we find evidence that the qubit's frequency stability produces a \SI{0.8}{\milli\second} limit on the pure dephasing which we also observe. These findings raise the need for performing qubit metrology to examine the reproducibility of qubit parameters, where these fluctuations could affect qubit gate fidelity. 

\end{abstract}

\maketitle
\section{Introduction}
Universal, fault-tolerant quantum computers---a Holy Grail of quantum information processing---are currently being pursued by academia and industry alike. To achieve fault tolerance in a quantum information processor, a scheme for quantum error correction\cite{Preskill} is needed due to the limited coherence lifetimes of its constituent qubits and the consequently imperfect quantum-gate fidelities. Such schemes, e.g.\@ the surface code\cite{FowlerSurface}, rely on gate fidelities exceeding a certain break-even threshold. Adequately high fidelity was recently demonstrated with superconducting qubits\cite{BarendsNat2014}; however, this break-even represents a best-case scenario without any temporal or device-to-device variation in the coherence times or gate fidelities. Therefore, a fault-tolerant quantum computer importantly requires not only improvements of the best-case single-\cite{Roltune} and two-qubit\cite{BarendsNat2014} gate fidelities: it actually requires the {\it typical} performance---in the presence of fluctuations---to exceed the error correction threshold. 
In the more immediate term, so-called Noisy Intermediate-Scale Quantum (NISQ)\cite{Perskillnisq} circuits will be operated without quantum error correction. In NISQ systems, gate fidelities and the fluctuations thereof directly limit the circuit depth, i.e.\@ the number of consecutive gates in an algorithm that can be successfully implemented.

In experiments with superconducting qubits, it is usual to perform qubit metrology\cite{Martinismetrology} to benchmark the gate fidelity and quantify its error, although these benchmarks are not typically repeated in time to determine any temporal dependence. Since gate fidelities are at least partially limited by qubit $T_1$ energy relaxation\cite{Roltune}, one would expect a fluctuation in gate fidelity resulting from a fluctuation in the underlying decoherence parameters. However, benchmarking of decoherence, to quantify the mean lifetime together with its stability or variation, is also uncommon. Consequently, it is unclear whether reports on improvements in coherence times---cf.\@ the review by Oliver and Welander\cite{Oliverreview} and that by Gu and Frisk Kockum et al.\cite{GU20171}---are reports of {\it typical} or of {\it exceptional} performance. Quantifying this difference is crucial for future work aimed at improving qubit coherence times. 

In this paper, we benchmark the stability of decoherence properties of superconducting qubits: $T_1$, $T_2^*$ (free-induction decay), $T_{\phi}$ (pure dephasing), and $f_{01}$ (qubit frequency). This study is distinct from numerous studies that report on singular measurements of qubit lifetimes for different background conditions, such as temperature\cite{wang2018cavity} or magnetic flux\cite{Dunsworthloss,Yanflux}.  Some studies\cite{Yanflux,Dialloss,Rosenberg3D,Muellertls,Gustavssonqp,ChangTiN,Klimovfluctuations} examine repeated measurements of qubit lifetimes under static conditions. 
However, when discussing these examples, it is important to quantify both the number of counts and the total duration of the measurement. Here, the number of counts relates to the statistical confidence, while the total duration relates to the timescale of fluctuations to which the study is sensitive. 
Therefore, to confidently report on fluctuations relevant to the calibration period of a quantum processor (for example a few times a day for the IBM Q Experience\cite{ibmq}), we only discuss reports featuring both a large number of counts ($N\!>\!1000$) and a total duration exceeding 5 hours. 

The first study to satisfy these requirements for relaxation measurements was that of M\"uller et al.\cite{Muellertls}, which revealed that unstable near-resonant two-level-systems (TLS) can induce fluctuations in qubit $T_1$. They proposed a model in which the TLS produces a strongly peaked Lorentzian noise profile at the TLS frequency (which is near the qubit frequency). Under the separate model of interacting TLS\cite{Faorointeracting,Burnettnoise}, the frequency of this near-resonant TLS varies in time. Consequently, the qubit probes the different parts of the TLS-based Lorentzian noise profile, leading to variations in the qubit's $T_1$. Although the mechanism was clearly demonstrated, this work\cite{Muellertls} was unable to determine properties of the TLS such as switching rates or dwell times of specific TLS frequency positions. Follow-up work by Klimov et al.\cite{Klimovfluctuations} used a tuneable qubit to map the trajectories of individual TLS. These findings\cite{Klimovfluctuations} supported the interacting-TLS model and M\"uller's findings, and were able to clearly determine TLS switching rates as well as reveal additional diffusive motion of the TLS. 

We demonstrate that sufficient statistical analysis can reveal the TLS-based Lorentzian noise spectrum and allow for extraction of switching rates. Importantly, this method does not require a tuneable qubit or advanced reset protocols\cite{Geerlingsreset} and is therefore general to any qubit or setup. Furthermore, the lack of tuning results in a more frequency-stable qubit and consequently less dephasing. This enables us to go beyond the studies of M\"uller et al.\@ and Klimov et al.\@ by studying the qubit's frequency instabilities due to other noise sources, which reveals a $1/f$ frequency noise that is remarkably similar to interacting-TLS-induced $1/f$ capacitance noise found in superconducting resonators\cite{Burnettnoise,Graafsuppression}. This frequency instability produces a limit on pure dephasing which we observe through sequential inter-leaved measurements of qubit relaxation, dephasing, and frequency.

\section{Results}
\subsection{Description of the devices}

\begin{table}
\centering
\begin{tabular}{ p{2.5cm} | p{0.5cm} p{2cm}  p{2cm} }
  \hline
  \hline			
  Parameters & & qubit A & qubit B \\
  \hline
  $f_R$ & & 6.035~GHz & 5.540~GHz \\
  $f_{01}$ & & 4.437~GHz & 3.759~GHz \\
  $f_{12}-f_{01}$ & & \SI{-0.226}{\giga\hertz} & \SI{-0.278}{\giga\hertz} \\
  $E_{J}/h$ & & 13.42~GHz & 8.57~GHz \\
  $E_{c}/h$ & & 0.201~GHz & 0.235~GHz \\
  $E_{J}/E_{c}$ & & 66.67 & 36.54 \\
  $\epsilon/h$ & & \SI{-524}{\hertz} & \SI{-109}{\kilo\hertz} \\
  \hline
  \hline
\end{tabular}
\caption{Summary of device parameters. $f_R$ is the frequency of the readout resonator and $f_{01}$ that of the qubit's 01 transition. $f_{12}-f_{01}$ is the frequency difference between the qubit's 12 transition and 01 transition. $E_{J}$ is the qubit's Josephson energy, $E_{c}$ its charging energy, $h$ is Planck's constant, and $\epsilon$ its charge dispersion.
}
\label{qubittab1}
\end{table}

Our circuit is made of aluminium on silicon and consists of a single-junction Xmon-type transmon qubit\cite{Barendsxmon} capacitively coupled to a microwave readout resonator (see the Methods section \ref{sec:exp_details} for more details). The shunt capacitor and the absence of magnetic-flux tunability (absence of a SQUID) effectively decouple the qubit frequency from electrical charge and magnetic flux, reducing the sensitivity to these typical 1/$f$ noise sources\cite{Gustavssoncharge,Bylandernoise}. Although these qubits lack frequency tunability, they remain suitable for multi-qubit architectures using all-microwave-based two-qubit-gates\cite{McKayUniversal, Chowgate, Economouswipht, Paikrip}. The circuit is intentionally kept simple so that the decoherence is dominated by intrinsic mechanisms and not external ones in the experimental setup. Therefore, there are no individual qubit drive lines, nor any qubit-to-qubit couplings. Additionally, both the spectral linewidth of the resonator and the resonator-qubit coupling are kept small, such that photon emission into the resonator (Purcell effect) and dephasing induced by residual thermal population of the resonator are minimized\cite{Clerkthermal}. A detailed experimental setup together with all device parameters are found in the Methods and \cref{qubittab1}.

This study involves two qubits on separate chips which we name A and B. The main differences are their Josephson and charging energies and that the capacitor of qubit B was trenched to reduce the participation of dielectric loss\cite{Barendsloss}.

\begin{figure}
    \centering
    \includegraphics[width=1\columnwidth]{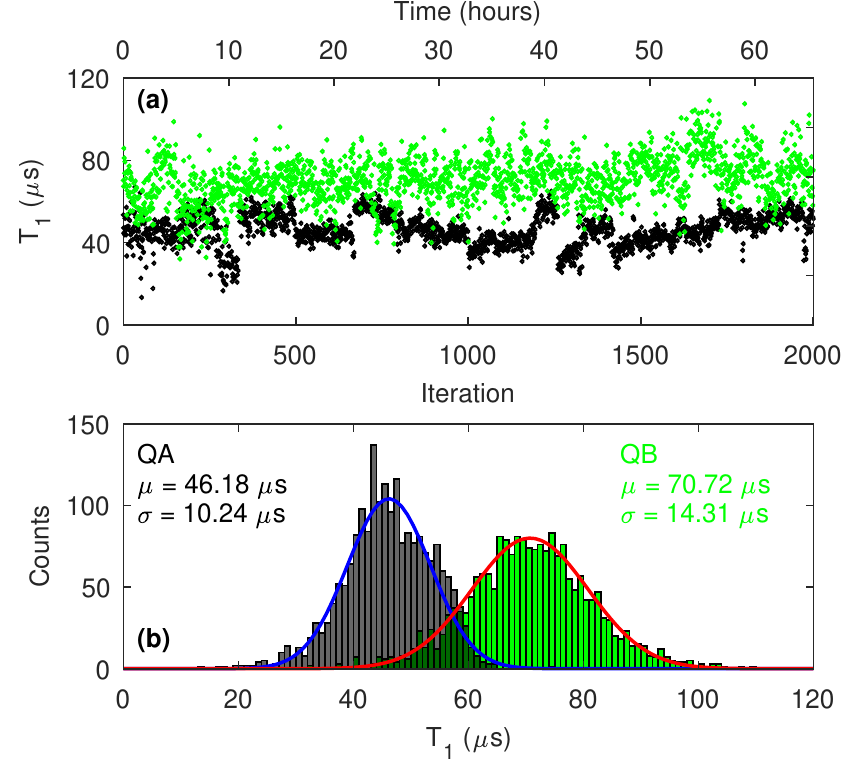}
    \caption{{\bf Synchronous measurement of $T_1$ in two separate qubits.} {\bf (a)} Multiple $T_1$ measurements performed simultaneously on qubits A (black) and B (green). The data consists of 2000 consecutive $T_1$ measurements that lasted a total duration of \SI{2.36d5}{\second} (approximately 65 hours). {\bf (b)} Histograms of the $T_1$ values in (a). The histograms have been fit (solid line) to Gaussian distributions with the parameters shown. This data was taken during cooldown 6. 
    }
    \label{fig:synced}
\end{figure}

\subsection{Synchronous \texorpdfstring{$T_1$}{T1} measurement of separate qubits}
First we assess the stability of the energy-relaxation time $T_1$ by consecutive measurements. The transmon is driven from its ground to first-excited state by a calibrated $\pi$ pulse. The qubit state is then read out with a variable delay. The population of the excited state, as a function of the readout delay, is fit to a single-exponential decay to determine $T_1$. \Cref{fig:synced} shows a 65-hour measurement of two separate qubits (in separate sample enclosures) that are measured simultaneously. The first observation is that the periods of low-$T_1$ values are not synchronized between the two qubits, indicating that the dominant mechanism for $T_1$ fluctuations is local to each qubit. (The lack of correlation is quantified in the Supplement.) In \cref{fig:synced}b, we histogram the $T_1$ data: this demonstrates that $T_1$ can vary by more than a factor 2 for both qubits, similarly to previous studies\cite{Muellertls,Klimovfluctuations}.

To make a fair comparison of the mean $T_1$ for two qubits with different frequencies, we can rescale to quality factors ($Q=2\pi f_{01}T_1$). We see that qubit B ($Q=1.67\times 10^6$) has a higher quality factor compared to qubit A ($Q=1.29\times 10^6$). However, while the quality of qubit B is higher, qubit B has a lower ratio of Josephson to charging energy (see Table~\ref{qubittab1}), resulting in a larger sensitivity to charge noise and parity effects\cite{Risteparity}. Consequently, qubit B exhibits switching between two different transition frequencies, which was not suitable for later dephasing and frequency instability studies. Therefore, most of the paper focuses on qubit A.

\subsection{\texorpdfstring{$T_1$}{T1} decay-profiles}
\begin{figure}
    \centering
    \includegraphics[width=1\columnwidth]{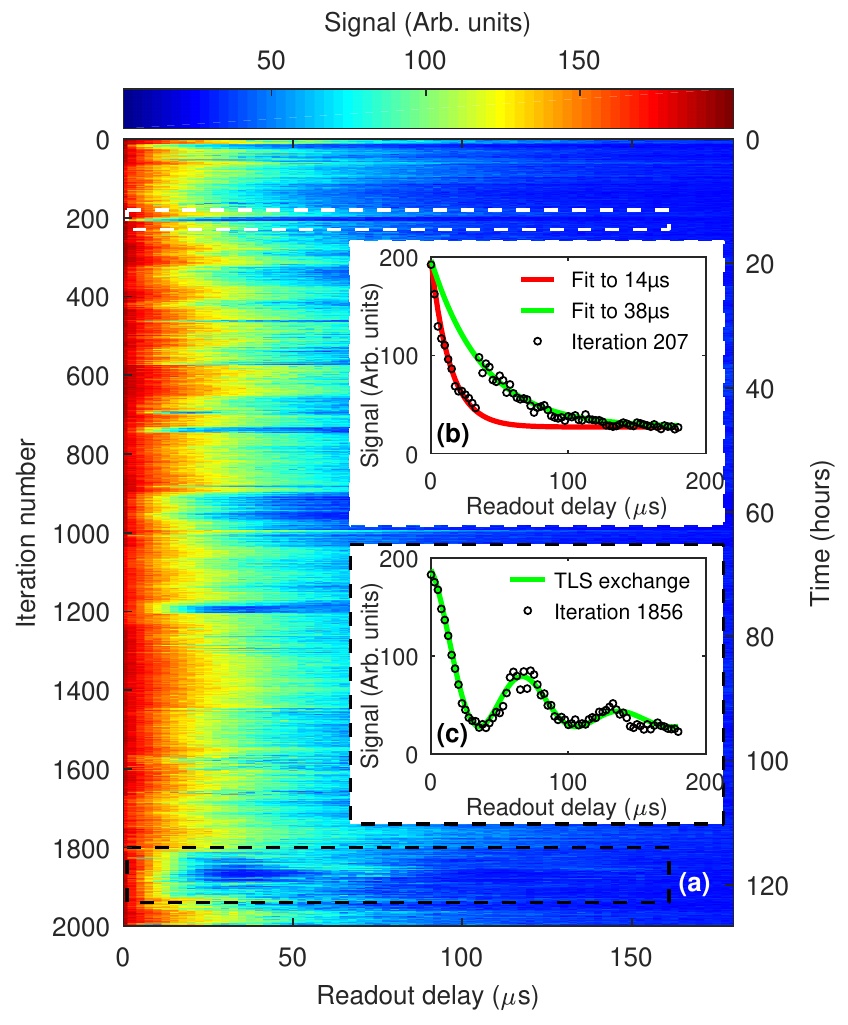}
    \caption{{\bf Raw data of $T_1$ decay-profile.} {\bf (a)} Consecutive $T_1$ measurements, spanning \SI{4.6d5}{\second} (approximately 128 hours), of qubit A. {\bf (b)} A data set showing a change in $T_1$ within a single iteration. These jumps are found to occur in approximately 3\% of all measurements. {\bf (c)} A data set showing a decaying sinusoidal (rather than a purely exponential) decay profile. The appearance of revivals are due to resonant exchange with a TLS. These profiles are found to occur in approximately 5\% of all iterations. This data was taken during cooldown 5.
    }
    \label{fig:T1shapes}
\end{figure}

We continue by measuring $T_1$ consecutively for approximately 128 hours, and plot the decays in a colour map (\cref{fig:T1shapes}a). Here, the colour map makes some features of the data simpler to visualize. Firstly, the fluctuations are comprised of a switching between different $T_1$ values, where the switching is instantaneous, but the dwell time at a particular value is typically between 2 and 12.5 hours. This behaviour (also seen in \cref{fig:synced}a) resembles telegraphic noise with switching rates ranging from \SI{20}{\micro\hertz} to \SI{140}{\micro\hertz}. Later, we quantify these rates and their reproducibility. 

\begin{figure}
    \centering
    \includegraphics[width=1\columnwidth]{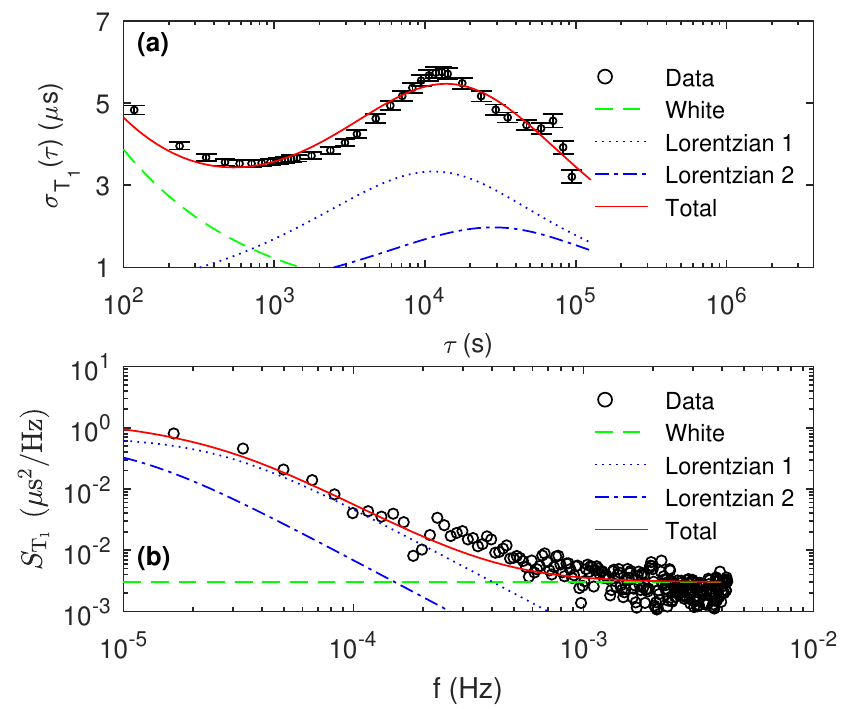}
    \caption{{\bf Time and frequency domain analysis of $T_1$ fluctuation.} Statistical analysis of 2001 sequential $T_1$ measurements of qubit A spanning a total measurement duration of \SI{2.36d5}{\second}. {\bf (a)} Overlapping Allan deviation of $T_1$ fluctuations.
    {\bf (b)} Welch-method spectral density of $T_1$ fluctuations. 
     In both plots there are fits to the total noise (red line) which is formed of white noise (green lines) and two different Lorentzians (blue lines). The amplitudes and time constants of all noise processes are the same for both types of analysis. This data was taken during cooldown 2. 
    }
    \label{fig:T1spectra}
\end{figure}

The white box of \cref{fig:T1shapes}a and inset \cref{fig:T1shapes}b show this switching behaviour occurring within a single iteration. The decay can be fit to two different values of $T_1$, one before the switch and one afterwards. This type of decay profile is found in approximately 3\% of the iterations. In all presented $T_1$ values (histograms or sequential plots), the lower $T_1$ value is used. This is motivated by quantum algorithms being limited by the shortest-lived qubit. 

The black box and inset \cref{fig:T1shapes}c highlight a decay-profile that is no longer purely exponential, but instead exhibits revivals. Similar revivals have been observed in both phase\cite{Cooperphasetls} and flux\cite{gustavssontls} qubits, and were attributed to coherently coupled TLS residing in one of the qubit junctions. From the oscillations we extract a qubit-to-TLS coupling of $g_{\mathrm{TLS}}=\SI{4.8}{\kilo\hertz}$. 
Assuming a TLS dipole moment of \SI{1}{\elementarycharge\angstrom}\cite{Martinisloss}, the coupling corresponds to an electric field line of \SI{39}{\micro\meter} (see the Supplement for more details). This length is larger than the Josephson junction; therefore, we conclude that this particular TLS is located on one of the surfaces of the shunting capacitor (not within the junction). Since the invention of transmons and improvement in capacitor dielectrics, individual TLS have only been found to incoherently couple to a transmon\cite{Barendsxmon}, and the authors are not familiar with any examples of a coherent coupling between a TLS and a transmon. 

Approximately 5\% of decay profiles show a clear revival structure, with a further 3\% showing hints of it. Of these, some revival shapes (such as the one shown in the black box) remained stable and persisted for approximately 10 hours, whereas others lasted for only 2--3 traces (around 10 minutes). Since the qubit here is fixed in frequency, these appearances/disappearances of the coherent TLS arise due to the TLS shifting in frequency\cite{Faorointeracting,Burnettnoise,Muellertls,Klimovfluctuations} relative to the static qubit. The observation of coherent oscillations in the decay, and in particular that oscillation periods remained stable for hours (for the same duration as the $T_1$ fluctuations), constitutes clear evidence for TLS being the origin of the $T_1$ fluctuations, in agreement with both the M\"uller\cite{Muellertls} and Klimov\cite{Klimovfluctuations} results. 

\subsection{Decoherence benchmarking}
To gain further insight into these fluctuations we perform statistical analysis commonly used in the field of frequency metrology. In parallel, we examine both the overlapping Allan deviation (\cref{fig:T1spectra}a) and the spectral properties (\cref{fig:T1spectra}b) of the $T_1$ fluctuations. Allan deviation is a standard tool for identifying different noise sources in e.g.\@ clocks and oscillators\cite{rubiola2009phase}. Here, we introduce the Allan deviation as a tool to identify the cause of fluctuations in qubits. The most striking feature in \cref{fig:T1spectra}a  is the peak and subsequent decay around $\tau$~=~10$^4$ seconds. Importantly, no power-law noise process can produce such a peak; instead, it is an unambiguous sign of a Lorentzian noise process. Such Lorentzian-like switching was observed in the $T_1$-vs.-time measurement in \cref{fig:synced}a. In \cref{fig:T1spectra}, we model the noise with two Lorentzians with a white noise floor, and apply the modelled noise to both the spectrum and the Allan deviation. Therefore, the noise parameters are the same for both plots: the Methods section has more details on the scaling of Lorentzian noise between the Allan and spectral analysis methods. From \cref{fig:T1spectra}, we obtain Lorentzian switching rates of  \SI{80.6}{\micro\hertz} and \SI{158.7}{\micro\hertz}.

\begin{figure*}
    \includegraphics[width=1.9\columnwidth]{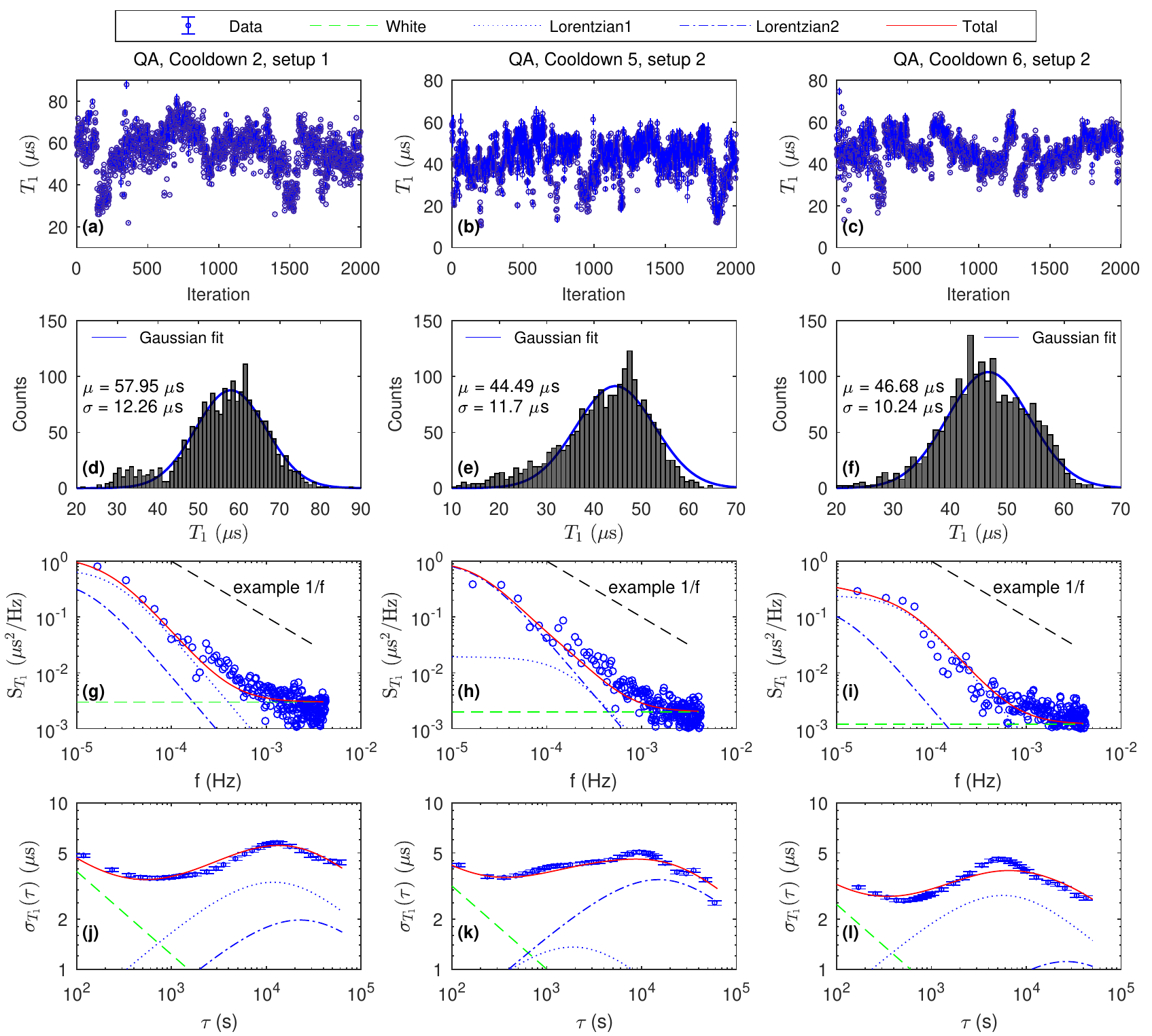}
    \caption{{\bf Reproducibility of $T_1$ fluctuations in qubit A across separate cooldowns.} ({\bf a-c}) Time evolution of $T_1$ vs.\@ iteration. ({\bf d-f}) Statistics of $T_1$ plotted as a histogram, with a Gaussian fit. ({\bf g-h})~Welch spectral density estimate of the $T_1$ fluctuations. ({\bf j-l}) Overlapping Allan deviation of $T_1$ fluctuations. Across ({\bf g-i}) and ({\bf j-l}) the noise model is the same, where the parameters can be found within \cref{noisetab}. For illustrative purposes, we include a $1/f$ noise guideline within ({\bf g-i}). Similar data for qubit B can be found within the Supplement.}
    \label{fig:combo}
\end{figure*}

Within \cref{fig:combo} and \cref{noisetab}, we show the reproducibility of these features across thermal cycles.  Collectively, we find switching rates ranging from \SI{71.4}{\micro\hertz} to \SI{1.9}{\milli\hertz}---slower than those obtained by measurements of charge noise\cite{Kafanovnoise} but similar to bulk-TLS dynamics\cite{Salvinodielectric,Ludwigdynamics} and in agreement with rates determined from measurements tracking the time evolution of individual TLS\cite{Klimovfluctuations}. These measurements demonstrate not only that superconducting qubits are useful probes of TLS, but unambiguously demonstrate the role of a TLS-based Lorentzian noise profile as a limiting factor to the temporal stability of qubit coherence. 

\begin{table}
\centering
\begin{tabular}{ p{1.15cm} | p{1.35cm} p{1.25cm} p{1.35cm} p{1.25cm} p{1.35cm}}
  \hline
  \hline
  & & & & & \\[-7pt]
  Data & $h_0$ (\si[per-mode=symbol]{\square \micro \second \per \hertz }) & $1/\tau_0^{Lor1}$ \; (\si{\micro\hertz}) & $A^{Lor1}$ (\si{\micro\second}) & $1/\tau_0^{Lor2}$ \; (\si{\micro\hertz}) & $A^{Lor2}$ (\si{\micro\second}) \\[1pt]
  \hline
  & & & & & \\[-7pt]
  QA\_C2 & 3.0$\times$10$^{-3}$ & 158.7 & 5.4 & 80.6 & 3.2\\[1pt]
    & & & & & \\[-7pt]
  QA\_C3 & 2.6$\times$10$^{-3}$ & 200.0 & 2.4 & 100.0 & 4.5\\[1pt]
      & & & & & \\[-7pt]
  QA\_C5 & 2.0$\times$10$^{-3}$ & 142.9 & 5.2 & 83.3 & 2.6\\[1pt]
        & & & & & \\[-7pt]
  QA\_C6 & 1.2$\times$10$^{-3}$ & 333.3 & 4.5 & 71.4 & 1.8\\[1pt]
          & & & & & \\[-7pt]
           \hline
               & & & & & \\[-7pt]
  QB\_C1 & 1.3$\times$10$^{-2}$ & 1851.8 & 2.5 & - & -\\[1pt]
      & & & & & \\[-7pt]
  QB\_C5 & 1.4$\times$10$^{-2}$ & 1000.0 & 3.2 & 90.9 & 6.6\\[1pt]
        & & & & & \\[-7pt]
  QB\_C6 & 5.7$\times$10$^{-3}$ & 1111.1 & 4.2 & 76.9 & 2.2\\[1pt]
  \hline
  \hline
\end{tabular}
\caption{Summary of the noise parameters for modeling $T_1$ fluctuations. The data is labeled as Q (qubit) A or B and C\# (\# denotes cooldown number). The superscripts \emph{Lor}1/2  correspond to the Lorentzian being parameterised.}
\label{noisetab}
\end{table}

\subsection{Interleaved measurements of \texorpdfstring{$T_1$}{T1}, \texorpdfstring{$T_2^*$}{T2*}, and \texorpdfstring{$f_{01}$}{f01}}
\begin{figure}
    \centering
    \includegraphics[width=0.95\columnwidth]{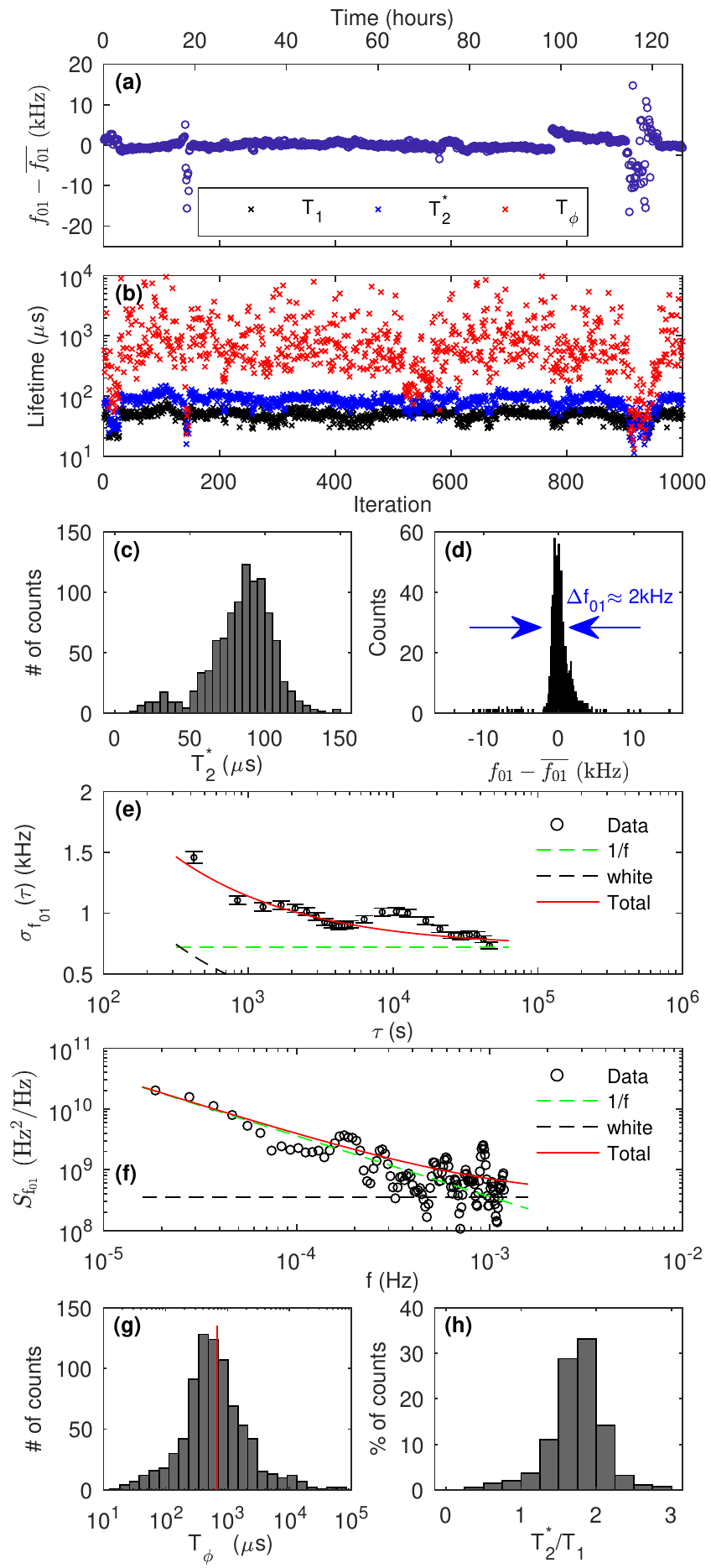}
    \caption{{\bf An interleaved series of 1000 $T_1$ relaxation and $T_2^*$ Ramsey measurements of qubit A.} {\bf (a)} Qubit frequency ($f_{01}$) shift relative to its mean ($\overline{f_{01}}$) determined from the Ramsey experiments. {\bf (b)} Extracted $T_1$ (black), $T_2^*$ (blue), and $T_\phi$ (red).  {\bf (c)} Histogram of $T_2^*$ from the data in (b). {\bf (d)} Histogram of the data in (a). The frequency fluctuations from (a) are analyzed by overlapping Allan deviation {\bf (e)} and by Welch-method spectrum {\bf (f)}. The solid and dashed lines represent the modeled noise, where the noise amplitudes are the same for both types of analysis. {\bf (g)} Histogram of $T_{\phi}$ from the data in (b). The solid line indicates the $T_{\phi}$ limit calculated by integrating the frequency noise from (e). {\bf (h)} Histogram of $T_2^*/T_1$ from the data in (b). We find $1.4<T_2^*/T_1<2.2$ in 81.7~\% of the counts. This data was taken during cooldown 3. 
    }
    \label{fig:coherence}
\end{figure}

In addition to studying $T_1$ fluctuations, we also explore fluctuations in qubit frequency and dephasing. To this end, we measure the qubit frequency and the characteristic decay time $T_2^*$ by means of a de-tuned Ramsey fringe. 
We interleave the Ramsey sequence, point-by-point, with the previously discussed $T_1$ relaxation sequence. For clarity, if we consider the energy-relaxation measurements in \cref{fig:T1shapes}, the main plot (a) represents the complete measurement set, which is formed from 2000 iterations. Each iteration (e.g.\@ either inset) consists of data points which are themselves the averaged results of 1000 repeated measurements. In the interleaved sequence, we measure the data point in the $T_1$ sequence and then the data point in the Ramsey sequence for each delay time (i.e.\@ the time between the $\pi$ pulse and readout, in the $T_1$ measurement, and in-between the $\pi/2$ pulses, in the Ramsey $T_2^*$ measurement). 
This  sequence is then looped through all values of the delay time to map out the $T_1$ and Ramsey decay profiles (i.e.\@ the iteration).
While averaging each point in the inner loop gives a longer iteration time, and increases the noise window to which the Ramsey fringe is sensitive\cite{Martinisnoise}, it allows for all qubit parameters to be known in each iteration. From the so-obtained $T_1$ and $T_2^*$ we calculate the pure-dephasing time $T_\phi$ from $1/T_2^*=1/2T_1+1/T_\phi$. 
These values are shown in Fig.~\ref{fig:coherence}b, and the histogram of $T_2^*$ values is shown in Fig.~\ref{fig:coherence}c.

In \cref{fig:coherence}a we have extracted, from the Ramsey fringes, the frequency motion of the qubit relative to its mean frequency ($f_{01}-\overline{f_{01}}$). In general, the observed frequency shifts are on the order of \SIrange{1}{3}{\kilo\hertz}, with infrequent shifts of up to \SI{20}{\kilo\hertz}. A histogram of the qubit frequency (\cref{fig:coherence}d) reveals a main peak with a full-width at half-maximum of approximately \SI{2}{\kilo\hertz}. These frequency shifts are significantly smaller than the approximately \SI{500}{\kilo\hertz} frequency instability found in flux-tuneable qubits\cite{Klimovfluctuations}. From the perspective of gate fidelity, a 1-kHz frequency shift should have negligible effect, meaning that our qubits are well suited for quantum information processing since no re-calibration of the qubit frequency is needed. However, a fluctuating qubit frequency necessarily leads to qubit dephasing so it is important to quantify this fluctuation and therefore aid in efforts to find, and mitigate, the noise source. 

To provide more information on possible mechanisms for the frequency instability, we examine both the overlapping Allan deviation (\cref{fig:coherence}e) and the spectrum of frequency fluctuations (\cref{fig:coherence}f). In red, the frequency noise is modelled to $A/f + B$, where the exponent of $f$ is 1. Similarly to the previous $T_1$ analysis, the noise model is scaled so that the red line has the same amplitude in both \cref{fig:coherence}e and \cref{fig:coherence}f. In this model, the $1/f$ noise amplitude is $A = \SI{3.6e5}{\square\hertz}$. 

\section{Discussion}

For both qubits, across all cooldowns, we found fluctuations in $T_1$ that could be described by Lorentzian noise with switching rates in the range from \SI{75}{\micro\hertz} to \SI{1}{\milli\hertz}. For all superconducting qubits, three relaxation channels are usually discussed: TLS, quasiparticles, and parasitic microwave modes. Of these, parasitic microwave modes should not cause fluctuation since they  are defined by the physical geometry. For quasiparticles in aluminium, we can compare our observed slow fluctuations with two quasiparticle mechanisms found in the qubit literature: the quasiparticle recombination rate is \SI{1}{\kilo\hertz}\cite{deVisserqpfluct}; the timescale of quasiparticle number fluctuations leads to rates in the range from \SIrange{0.1}{10}{\kilo\hertz}; and finally quasiparticle tunnelling (parity switching events) in transmons have rates in the range \SIrange{0.1}{30}{\kilo\hertz}\cite{Risteparity}. Therefore, fluctuations in the properties of the superconductor occur over rates which differ by over six orders of magnitude compared to those found in our experiment. Instead, we highlight that, at low temperatures, bulk-TLS dynamics\cite{Salvinodielectric,Ludwigdynamics} and TLS-charge noise\cite{Kafanovnoise,gustavssontls} vary over long timescales equivalent to rates in the range from  \SI{10}{\milli\hertz} to \SI{100}{\hertz}. 

The observed coherent qubit--TLS coupling (\cref{fig:T1shapes}c) is an unambiguous sign of the existence of near-resonant TLS. Its fluctuation follows similar time constants as the $T_1$ fluctuations, which constitutes clear evidence of spectral instability, as expected from the interacting-TLS model\cite{Faorointeracting,Burnettnoise}. We therefore attribute the origin of the $T_1$ decay to near-resonant TLS, 
and the Lorentzian fluctuations in the qubit's $T_1$ (shown in \cref{fig:T1spectra} and \cref{fig:combo}g-l) arise due to spectral instabilities of the TLS  as described by M\"uller et al.\cite{Muellertls}. The extracted switching rates then represent the rate at which the near-resonant TLS is changing frequency. Similarly, the quality factor of superconducting resonators has also been found to vary\cite{earnest2018substrate} due to spectrally unstable TLS.

In general, we find that two separate Lorentzians are required to describe the fluctuation. This does not necessarily imply the existence of two near-resonant TLS---instead it is a limitation of the analysis, as we cannot resolve the difference between, say, two near-resonant TLS, each with two preferential frequencies, vs.\@ one near-resonant TLS that has four preferential frequencies. Such a difference could be inferred by measuring the local density of near-resonant TLS\cite{Klimovfluctuations}, although such a measurement has demonstrated that both scenarios above are possible\cite{Klimovfluctuations}. Additionally, when repeating the measurements across multiple cooldowns, we consistently find a near-resonant TLS that follows similar switching statistics. Between each cooldown, the TLS configuration is expected to completely change. Essentially, this means that the detuning and coupling of the observed near-resonant TLS should vary for each cooldown. However, despite any expected reconfiguration, at least one spectrally-unstable near-resonant TLS is always found to exist.


When examining the frequency stability of qubit A, we found a frequency noise of approximately \SI{2}{\kilo\hertz}, which was well described by a $1/f$ amplitude of \SI{3.6e5}{\square\hertz} (\cref{fig:coherence}d-f). Typically, dephasing is thought to arise due to excess photons within the cavity\cite{Rigettit2,wang2018cavity}, flux noise\cite{Bylandernoise}, charge noise\cite{gustavssontls,Kafanovnoise}, quasiparticles tunnelling through the Josephson junctions\cite{Risteparity}, or the presence of excess quasiparticles\cite{Barendsqp}. For qubit A, the charge dispersion is calculated to \SI{524}{\hertz}, much smaller than most of the observed frequency shifts. This rules out charge noise and tunnelling quasiparticles as the main source of the observed frequency fluctuations. Quasiparticle fluctuations have been extensively studied\cite{deVisserqpfluct}, where the magnitude of frequency shifts scales with the kinetic inductance. Therefore, while they can be of order \SI{100}{\kilo\hertz} in disordered superconductors\cite{grunhaupt2018quasiparticle}, they are much smaller in elemental superconductors. In fact, recent experiments\cite{deVisserqpfluct} showed that the quasiparticles in aluminium produced an un-measurably small frequency shift; instead, the quasiparticles' influence was revealed only by examining correlated amplitude and frequency noise. Therefore, not only do quasiparticles produce immeasurably small frequency shifts, but, as noted earlier, they act over much shorter timescales (i.e. rates are equivalent to kHz\cite{deVisserqpfluct,Risteparity} rather than the \si{\micro\hertz} observed here).

Instead we highlight two further TLS-based mechanisms. Firstly, TLS within the Josephson junction can cause critical-current noise\cite{nugrohonoise}, which can produce a frequency noise by modulating the Josephson energy. 
Alternatively, superconducting resonators demonstrate that TLS can produce frequency instabilities\cite{Burnettnoise,Graafsuppression} (capacitance noise). Both of these mechanisms exhibit a $1/f$ noise, where the noise amplitude is close to that which we find here.
One could distinguish between these two effects by examining the temperature dependence of the qubit's frequency noise. Here, critical current noise\cite{nugrohonoise} scales $\propto T$ while capacitance noise\cite{Burnettnoise,Graafsuppression} scales $\propto 1/T^{1.3}$.

Irrespective of the origin of the frequency instability, the noise spectrum in \cref{fig:coherence}f can be integrated to estimate the pure dephasing of the qubit\cite{Kochtransmon}. From this calculation, the expected $T_{\phi}$ is \SI{0.8}{\milli\second}. In \cref{fig:coherence}g we histogram the $T_{\phi}$ to reveal a peak around \SI{0.7}{\milli\second}, with diminishing counts above \SI{1}{\milli\second}, in good agreement with the estimate from the integrated frequency noise. 

In \cref{fig:coherence}b and \cref{fig:coherence}c, $T_2^*$ is almost always longer than $T_1$, implying that $T_{\phi} > 2T_1$. In \cref{fig:coherence}h this is quantified, as the histogram of the ratio of $T_2^*/T_1$ reveals that the qubit dephasing is almost always near 2$T_1$. Therefore, the qubit's $T_2^*$ is mainly limited by $T_1$. To the authors' knowledge all other demonstrations of $T_1$-limited $T_2$ required dynamical decoupling by either a  Hahn-echo (spin-echo)\cite{Bylandernoise,wang2018cavity} or CPMG\cite{Yanflux} sequence. However, neither of those works provide any statistics on whether the qubits were {\it always} $T_1$-limited. The histogram in \cref{fig:coherence}h also reveals counts where the ratio is above 2: these correspond to the instances where the $T_1$ has fluctuated within an iteration, similar to that shown in \cref{fig:T1shapes}b.

In summary, we have measured the stability of qubit lifetimes across more cooldowns and for measurement spans longer than previous studies. Collectively, this demonstrates that qubit fluctuations, due to spectrally unstable TLS, are consistently observed, even when $T_1$ is high (approaching \SI{100}{\micro\second}). Consequently, this demonstrates why it is necessary to re-calibrate qubits every few hours. Fundamentally, this also demonstrates that future reports on qubit coherence times require not only statistics for reproducibility, but also that the measurement duration should exceed several hours in order to adequately report the {\it typical} rather than {\it exceptional} coherence time. 

{\it Note added} --- Recently, a preprint on comparable observations was published by Schl\"or et al.\cite{schlortls}, who independently demonstrated that single fluctuators (TLS) are responsible for frequency and dephasing fluctuations in superconducting qubits. Additionally, another recent preprint by Hong et al.\cite{Honggate}, specifically measures fluctuations in gate fidelity and independently identifies $T_1$ fluctuation of the underlying qubits as the probable cause. 

\section{Methods} \label{sec:methods}

\subsection{Experimental details} \label{sec:exp_details}

\begin{figure}[H]
    \centering
    \includegraphics[width=0.85\columnwidth]{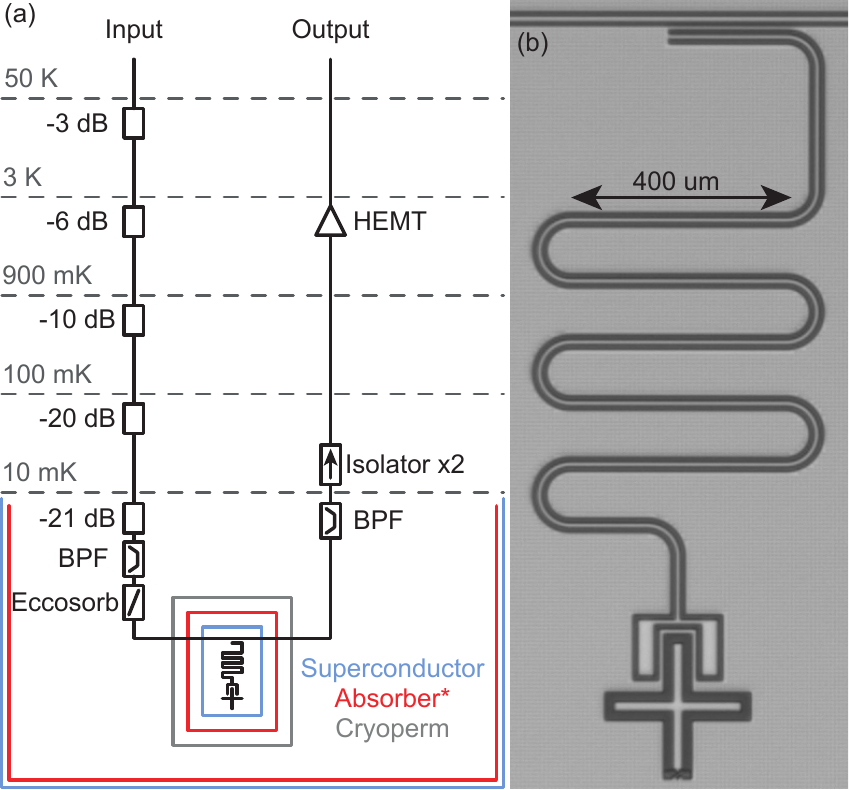}
    \caption{{\bf An overview of the experiment.}  ({\bf a}) Simplified schematic of the experimental setup. The main features are the various shielding layers. The absorber coating (red and with an asterisk) was added for setup 2. ({\bf b}) Optical image of the qubit sample. It shows a common microwave transmission line, a $\lambda/4$ resonator and a transmon qubit with a coplanar capacitor (Xmon-geometry).}
    \label{fig:setup}
\end{figure}

The qubits are fabricated out of electron-beam evaporated aluminium on a high-resistivity intrinsic silicon substrate. Everything except the Josephson junction is defined using direct-write laser lithography and etched using wet chemistry. The Josephson junction is defined in a bi-layer resist stack using electron-beam lithography, and later deposited using a two-angle evaporation technique that does not create any extra junctions or floating islands\cite{GladchenkoManhattan}. An additional lithography step is included to ensure a superconducting contact between the junction and the rest of the circuit; after the lithography, but prior to deposition of aluminium, an argon ion mill is used to remove native aluminium oxide. This avoids milling underneath the junction, which has been shown to increase the density of TLS\cite{Dunsworthloss}. Finally, the wafer is diced into individual chips and cleaned thoroughly using both wet and dry chemistry. Moreover, qubit B underwent a trenching step where approximately \SI{1}{\micro\meter} of the silicon dielectric was removed from both the qubit and the resonator using an fluorine based reactive-ion etch\cite{BurnettLT}.

A simplified schematic of the experimental setup is shown in \cref{fig:setup}a. The samples sit within a superconducting enclosure, which itself is inside of an absorber-lined radiation shield  and a cryoperm layer. This is located within a further absorber-lined radiation shield and a further superconducting layer which encloses the entire mixing chamber. Everything inside the cryoperm layer (screws, sample enclosures, and cables) is non-magnetic. The setup, including absorber recipe, is similar to a typical qubit box-in-a-box setup\cite{Barendsqp}. For the different cooldowns, two setups (labeled 1 and 2) were used. Setup~2 was as described above, whereas setup~1 lacked the absorber coating marked with a red asterisk in \cref{fig:setup}a.

\subsection{Data handling}

The qubit decoherence data is processed in the following way. First, the digitizer signal is rotated to one quadrature. Next, the signal is normalized to the maximum visibility of the qubit $\left|0\right>$ and $\left|1\right>$ states. Then, for qubit relaxation data, a fit to a single exponential is performed. Within \cref{fig:combo}a--c and the supplemental \cref{fig:combob}a--c, $T_1$ data is presented with error-bars. These error-bars correspond to 1 standard deviation, determined from confidence intervals of the exponential fit. For the Ramsey measurements, the initial processing is as described above. However, the Ramsey frequency ($f_{Ram}$) is initially determined by FFT of the data. The resulting frequency from the FFT is used as an initial frequency guess to a model of the form:
\begin{equation}
   \left<P_{e}(t)\right> = \exp(-t/T_{2}^*)\cos({\rm 2}\pi{f_{Ram}} + \phi_0)
    \label{T2Ramsey}
\end{equation}
where $\phi_0$ is a phase offset that is generally zero. Across all of the data-sets examined for qubit A, the FFT reveals only one oscillation frequency, whereas for qubit B, two frequencies are observed due to a larger charge dispersion. Consequently, \cref{T2Ramsey} doesn't fit well for qubit B, and we omit qubit B from the dephasing and frequency results.

For the qubit relaxation, we did attempt fits to a double exponential model\cite{popqpsuppression,Yanflux}. Within this model, an additional relaxation channel due to quasiparticles near the junction can lead to a skewing of the decay-profile. Here, we found the confidence interval for numerous parameters was un-physically large, indicating that the model over-parameterized our data. Therefore, we continued to use the single-exponential model. However, this is not surprising as the double-exponential is typically used for flux-qubits and fluxoniums, rather than the single-junction transmon-type qubit studied here.

\subsection{Sample handling}
Here, we clarify the sample handling across the entire experiment. For each qubit sample, after completing fabrication, they were covered in protective resist until the morning of their first cooldown (cooldown 1 for qubit B and cooldown 2 for qubit A). After removal of the resist, the samples were wire-bonded within a sample enclosure. Once sealed, the samples remained within their enclosures and were kept attached to the fridge for the entire experimental run. Therefore, when the fridge was warm, the samples were kept at the ambient conditions of the lab. Qubit B was not measured between cooldown 2 and cooldown 5, although it was still cooled down. However, qubit B was examined again in cooldowns 5 and 6 to gather statistics on the reproducibility of parameter fluctuations. 

\subsection{Spectral and Allan analysis}

Within the main text, information on TLS switching rates is inferred by examining the reproducibility of coherence parameters. Primarily, this is obtained by examining the Allan statistics and spectral properties of $T_1$ fluctuations. Here, the same data set is used to produce a plot of a Welch-method FFT ($S_{T_1}(f)$) and an overlapping Allan deviation ($\sigma_{T_{1}}(\tau)$). For the Welch analysis, the quantity analysed is $T_1 - \overline{T_1}$ (or for frequency is is $f_{01} - \overline{f_{01}}$). Therefore, the analysed quantity is not presented in fractional units. Consequently, the units of spectral analysis are \si[per-mode=symbol]{\square\micro\second \per \hertz} for $T_1$ fluctuations (Hz$^2$/Hz for frequency analysis). Equivalently, the Allan statistics are presented in units of \si{\micro\second} for $T_1$ fluctuation (Hz for frequency fluctuation).

The Allan deviation offers a few advantages compared to the spectrum. The method is directly traceable in that the Allan methods use simple mathematical functions that do not require any careful handling of window functions or overlap. When examining low-frequency processes, this eases a considerable burden in FFT-analysis which is to distinguish real features from remnants of window functions. This traceability is core to the usage within the frequency metrology community. The Allan method also provides clear error bars (defined as equal to 1 standard deviation), which translate to an efficient use of the data with optimum averaging of all data that shares a common separation, that is, all data pairs for any separation ($\tau$ in the Allan plot) are averaged over. Moreover, the Allan method can distinguish linear drift from any other divergent noise processes. Within an FFT, a linear drift appears as a general $1/f^a$ slope where $a$ is not unique compared to other noise sources. Within the Allan, a linear drift appears as $\tau^a$ where $a$ is distinct and unique compared to other divergent noise types.

From here, beginning with the Allan deviation, we consider the standard power-law model\cite{rubiola2009phase} of noise processes, 
\begin{equation}
\begin{aligned}
   \sigma_{T_1}(\tau) = {} & \left(\frac{h_{0}}{2}\right)^{\frac{1}{2}}\tau^{-\frac{1}{2}} \\
   & + \left(2{\rm ln}(2)h_{-1}\right)^{\frac{1}{2}} \\
   & + \left(\frac{(2\pi)^2}{6}h_{-2}\right)^{\frac{1}{2}}\tau^{\frac{1}{2}}
    \label{Allanpowlaw}
    \end{aligned}
\end{equation}
which can also be represented as spectral noise
\begin{equation}
S_{T_1}(f) = \frac{h_{-2}}{f^{2}} + \frac{h_{-1}}{f} + h_{0}
    \label{Welchpowlaw}
\end{equation}
where, in frequency metrology notation, $h_{-2}$ is the amplitude of a random walk noise process, $h_{-1}$ is the amplitude of a $1/f$ noise process and $h_0$ is the amplitude of 
white noise. 

In general terms, the power-law noise processes create a well-like shape in the Allan analysis, where, with the terms listed above, the walls have slopes of $\propto \tau^{-\frac{1}{2}}$ and $\propto \tau^{\frac{1}{2}}$. If more terms are included in the power-law noise model, the available slope gradients increase, but the well-like shape remains. When applied to the $T_1$ fluctuations (\cref{fig:T1spectra}), this model is not able to describe the most striking feature: the hill-like peak with subsequent second decreasing slope. Within Allan analysis, the rise and fall of a single peak can only be represented by a Lorentzian noise process. Therefore, starting from 
\begin{equation}
S(f) = \frac{4A^{2}\tau_{0}}{1 + (2\pi{f}\tau_{0})^2}
    \label{Welchlorentzian}
\end{equation}
where $A$ represents the Lorentzian noise amplitude and $\tau_{0}$ is the characteristic timescale, Lorentzian noise can be represented in Allan deviation by\cite{van1982new}:
\begin{equation}
\sigma(\tau) = \frac{A\tau_{0}}{\tau}\left(4e^{(-\tau/\tau_0)} - e^{(-2\tau/\tau_0)} -3 + 2\tau/\tau_0\right)^{1/2}
    \label{Allanlorentzian}
\end{equation}

From here, we model the $T_1$ fluctuations by two separate Lorentzians and white noise. When plotted, the noise from these sources is identical (i.e.\@ the same $h_0$, $A$, and $\tau_0$) for both the Welch-FFT and Allan deviation. For the rest of the data sets, we tabulate the Lorentzian parameters and white noise level in \cref{noisetab}.

\begin{acknowledgments}
We wish to express our gratitude to Philip Krantz and Tobias Lindstr\"om for insightful discussions. We acknowledge financial support from the Knut and Alice Wallenberg Foundation, the Swedish Research Council, and the EU Flagship on Quantum Technology H2020-FETFLAG-2018-03 project 820363 OpenSuperQ.
\end{acknowledgments}

\subsection{Author contributions}

J.J.B and A.B are considered co-first authors. J.J.B., A.B. and J.B planned the experiment, A.B. designed the samples, A.B fabricated the samples with input from J.J.B. The experiments were mainly performed by J.J.B and A.B with help from M.S, D.N and M.K. Analysis was performed by J.J.B and A.B with input from P.D and J.B. The manuscript was written by J.J.B, A.B and J.B with input from all authors. J.B and P.D provided support for the work. 

\subsection{Competing interests}

The authors declare that there are no competing interests. 

\subsection{Data availability}

The data that supports the findings of this study is available from the corresponding authors upon reasonable request.

\subsection{Code availability}

The code that supports the findings of this study is available from the corresponding authors upon reasonable request.

\subsection{Corresponding author}

Jonathan~Burnett (jonathan.burnett@npl.co.uk) or Jonas~Bylander (bylander@chalmers.se)

\bibliographystyle{naturemag}
\bibliography{bib.bib}

\clearpage

\section{Supplemental}

\subsection{Reproducibility of data}

\begin{figure}
    \includegraphics[width=0.9\columnwidth]{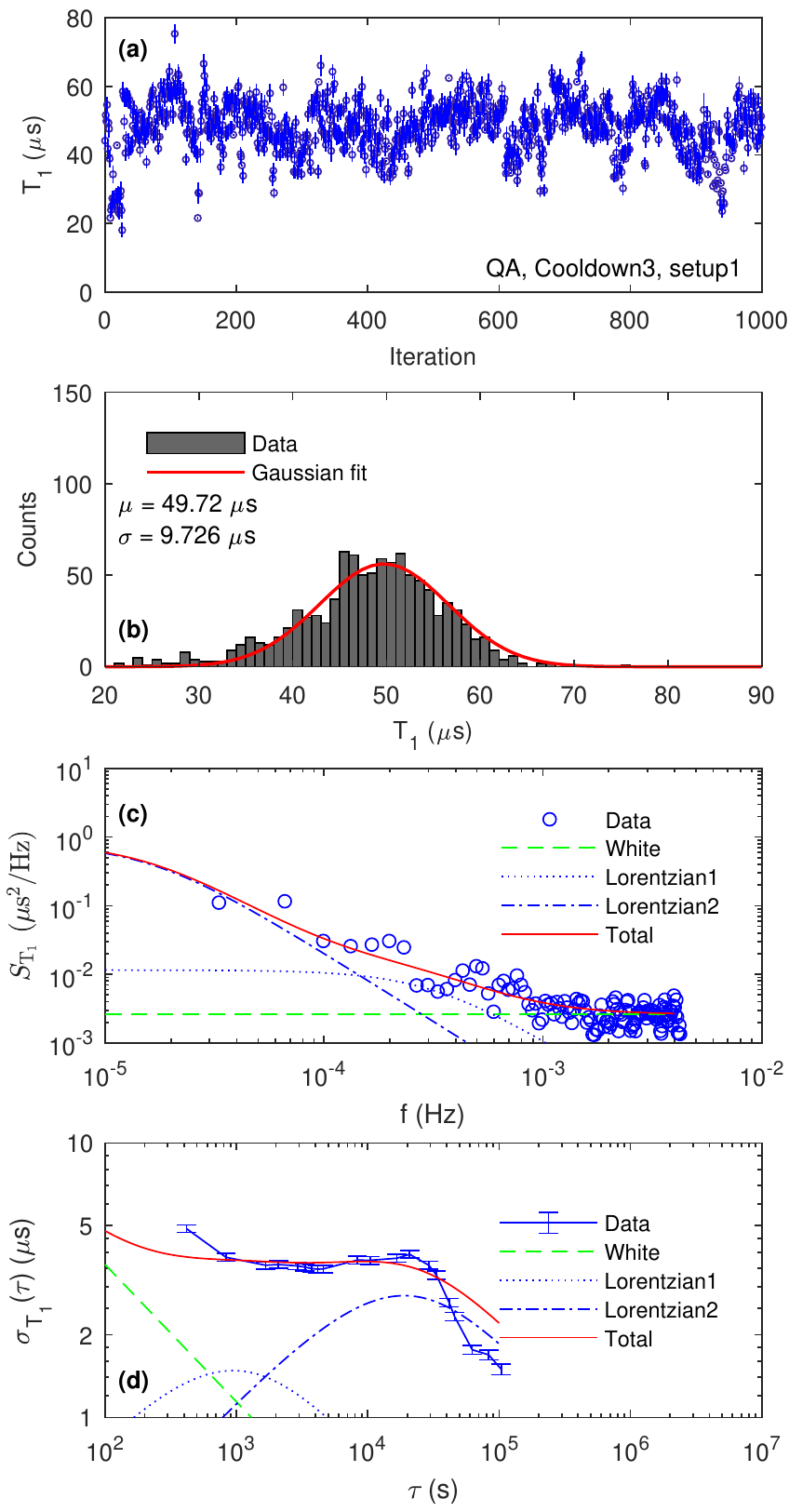}
    \caption{{\bf 1000 sequential $T_1$ measurements of Qubit A within cooldown 3.} ({\bf a}) $T_1$ vs.\@ iteration. ({\bf b}) $T_1$ histogram. ({\bf c})~Welch spectral density estimate of the $T_1$ fluctuations. ({\bf e}) Overlapping Allan deviation of $T_1$ fluctuations.}
    \label{fig:c3}
\end{figure}

\begin{figure*}
    \includegraphics[width=1.95\columnwidth]{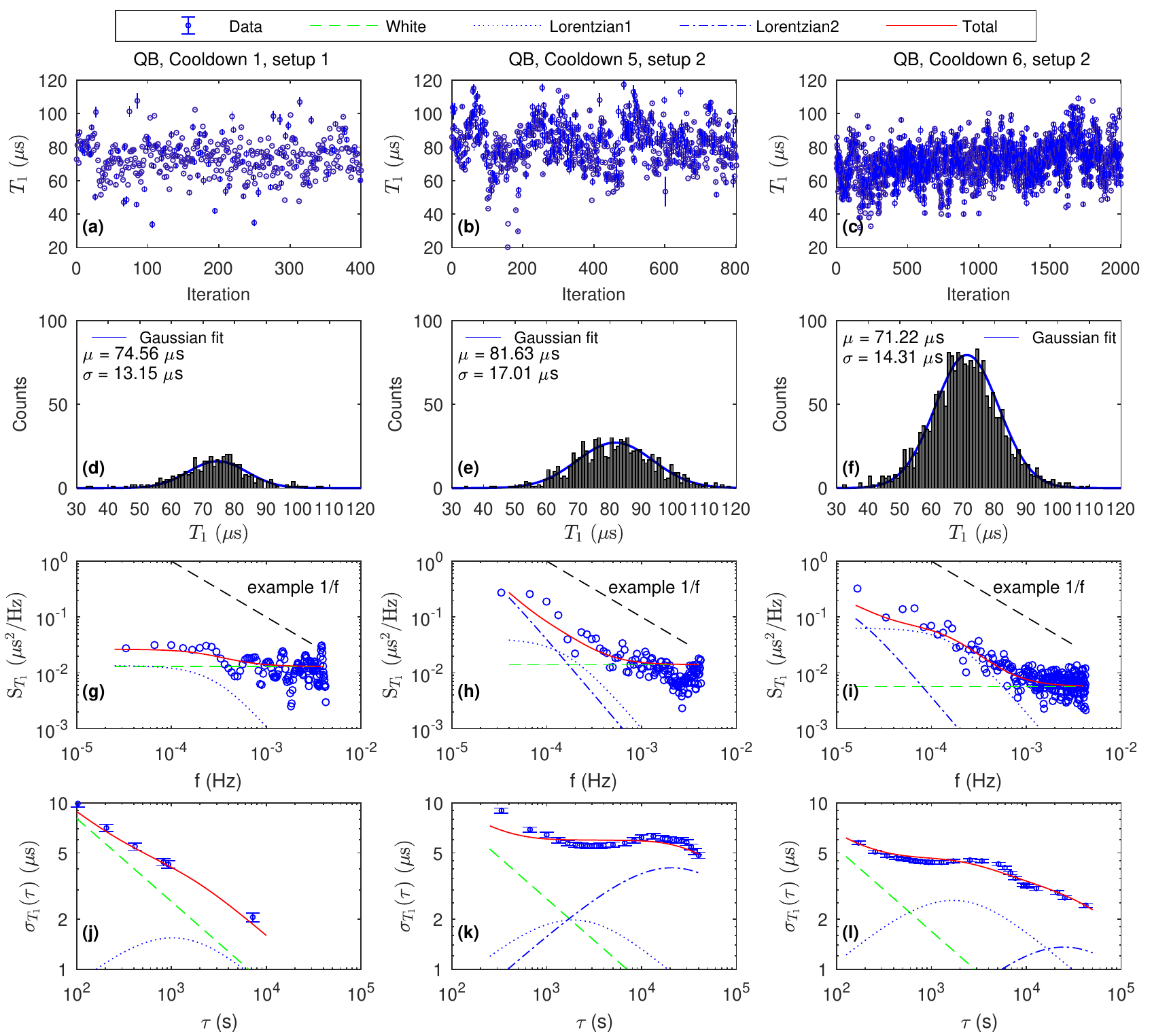}
    \caption{{\bf Reproducibility of $T_1$ fluctuations in qubit B across separate cooldowns.} ({\bf a-c}) Time evolution of $T_1$ vs.\@ iteration. ({\bf d-f}) Statistics of $T_1$ plotted as a histogram, with a Gaussian fit. ({\bf g-h})~Welch spectral density estimate of the $T_1$ fluctuations. ({\bf j-l}) Overlapping Allan deviation of $T_1$ fluctuations. Across ({\bf g-i}) and ({\bf j-l}) the noise model is the same, where the parameters can be found within \cref{noisetab}. For illustrative purposes, we include a $1/f$ noise guideline within ({\bf g-i}).}
    \label{fig:combob}
\end{figure*}

In the main text, the reproducibility of data for qubit A was shown by summaries of $T_1$ values vs.\@ measurement iteration; a histogram of $T_1$ values with a fit determining the mean and standard deviation; and a Welch FFT and overlapping Allan deviation of the $T_1$ series. That summary only featured the datasets with the highest number of counts. Therefore, the remaining data set is included here (see \cref{fig:c3}). Additionally we include the data sets for qubit B (se \cref{fig:combob})

\begin{figure}
    \centering
    \includegraphics[width=0.9\columnwidth]{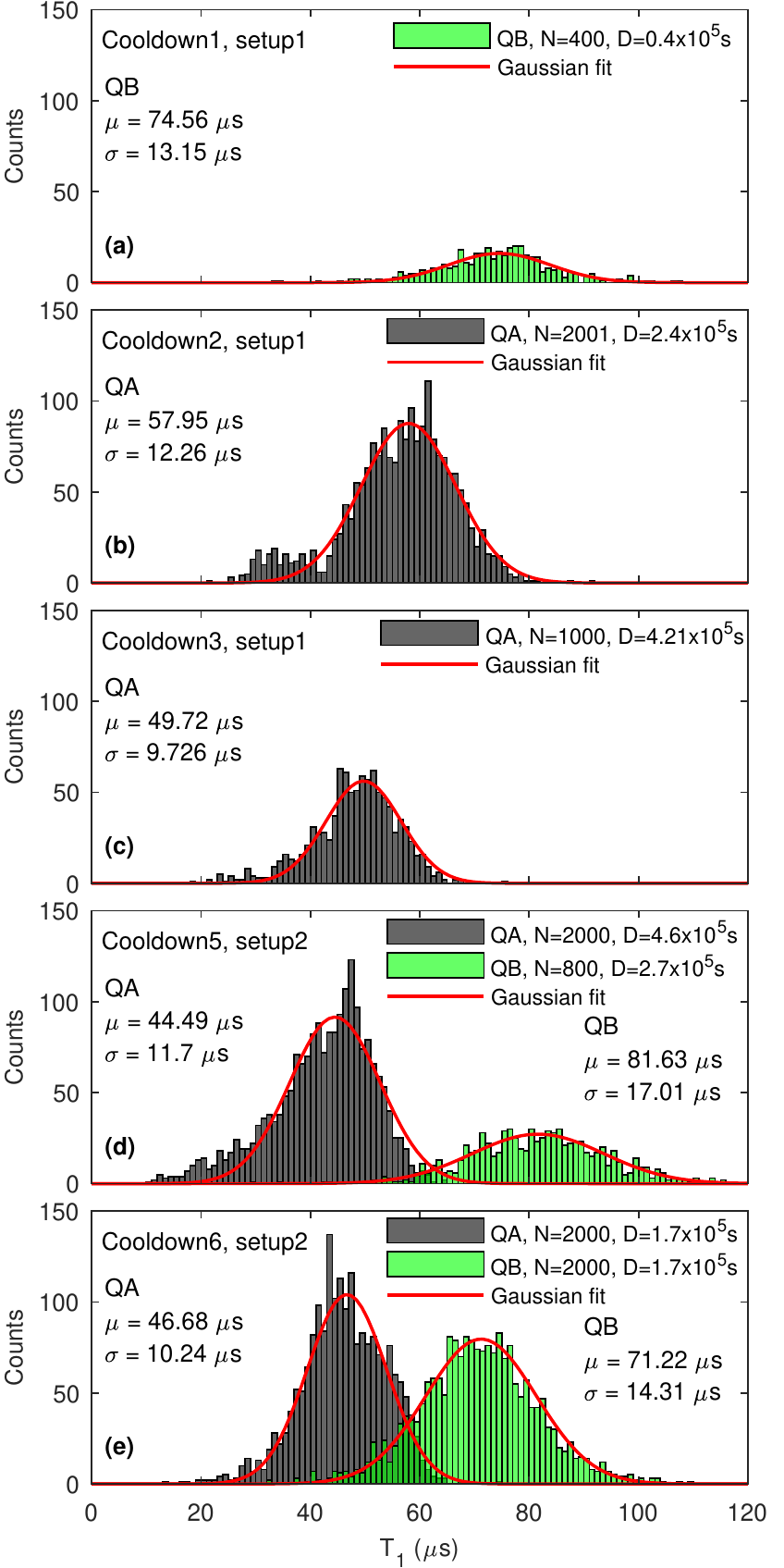}
    \caption{{\bf Statistics on $T_1$ across cooldowns.} A series of histograms from $N$ measurements spanning a measurement duration ($D$) for qubit~A (in black) and qubit~B (in green) across several separate cooldowns. Setup~2 represents the full schematic demonstrated in \cref{fig:setup}a, while setup~1 does not include the absorber coating from \cref{fig:setup}a}
    \label{fig:aging}
\end{figure}

\begin{table}[b]
\centering
\begin{tabular}{ p{1.4cm} | p{1.cm} p{1.cm} p{1.cm} | p{1.cm} p{1.cm} p{1.cm}}
  \hline
  \hline
  & & & & & & \\[-7pt]
  & QA & & & QB & &\\
  Cooldown & $\overline{T_1}$ (\si{\micro\second}) & days$_{c}^{tot}$ & days$_{w}^{tot}$ & $\overline{T_1}$ (\si{\micro\second}) & days$_{c}^{tot}$ & days$_{w}^{tot}$ \\[1pt]
  \hline
  & & & & & & \\[-7pt]
  1 & - & - & - & 74.56 & 0 & 1 \\[1pt]
  2 & 57.95 & 0 & 1 & - & 39 & 5 \\[1pt]
  3 & 49.72 & 49 & 5 & - & 88 & 9 \\[1pt]
  4 & - & 80 & 10 & - & 119 & 14 \\[1pt]
  5 & 44.49 & 133 & 15 & 81.63 & 192 & 19 \\[1pt]
  6 & 46.68 & 197 & 23 & 71.22 & 236 & 27 \\[1pt]
  \hline
  \hline
\end{tabular}
\caption{Statistics on qubit $T_1$ over time. We tabulate the mean $T_1$ (denoted $\overline{T_1}$), the cumulative total days cold (days$_{c}^{tot}$), and the cumulative total days warm (days$_{w}^{tot}$).}
\label{agetab}
\end{table}

By examining the reproducibility of $T_1$ fluctuations across thermal cycles, the study grew to span multiple months. Between these thermal cycles, the samples were kept at ambient conditions within the laboratory room. Therefore, with the many thermal cycles, the samples may spend a non-negligible amount of time outside of the controlled vacuum and cryogenic conditions of the cryostat. Therefore, we start to raise the possibility of becoming sensitive to the device aging. Aging of Josephson junctions is well-known and often reported as causing a drift in the Josephson energy. 

Here, we are interested in whether the qubit lifetimes degrade over time. Anecdotally, there is an awareness that devices age, where, for example, increased surface oxidation can increase dielectric losses. However, the authors are not aware of any studies which demonstrate robust statistics on any aging process. In \cref{fig:aging}, we show several measurements of $T_1$ statistics across several cooldowns. These demonstrate that the observed fluctuations are {\it typical} for all cooldowns. Additionally, we find that the standard deviation of the $T_1$ fluctuation is around 20\% of the mean, for mean $T_1$ ranging from \SI{40}{\micro\second} to \SI{82}{\micro\second}. 

\Cref{agetab} quantifies the number of days that the samples were cold (temperature below \SI{3}{\kelvin}) or at ambient conditions. We observe that qubit A experiences degradation within the first 15 days at ambient conditions. Within this time, the mean $T_1$ drops from $\approx$~\SI{60}{\micro\second} to around $\approx$~\SI{45}{\micro\second}, where no further degradation is observed. Here, we are limited to too few samples to make definitive statements on qubit aging. None-the-less, the observation of degradation across the cooldowns is a hint that aging could occur.  Importantly, we demonstrate that in order to actually resolve the small degradation in performance, large statistics (larger than typically reported) are essential.

\subsection{Fits to alternative noise models}

In the main text, we have modeled the $T_1$ fluctuations by two different Lorentzians. Primarily, this was motivated by the peak within the Allan deviation that could not be understood by other noise processes (e.g.\@ $1/f$ noise). However, there are some sets of data where the Lorentzian peak is not that prominent (e.g. Cooldown 5 for either qubit).

\begin{figure*}
    \centering
    \includegraphics{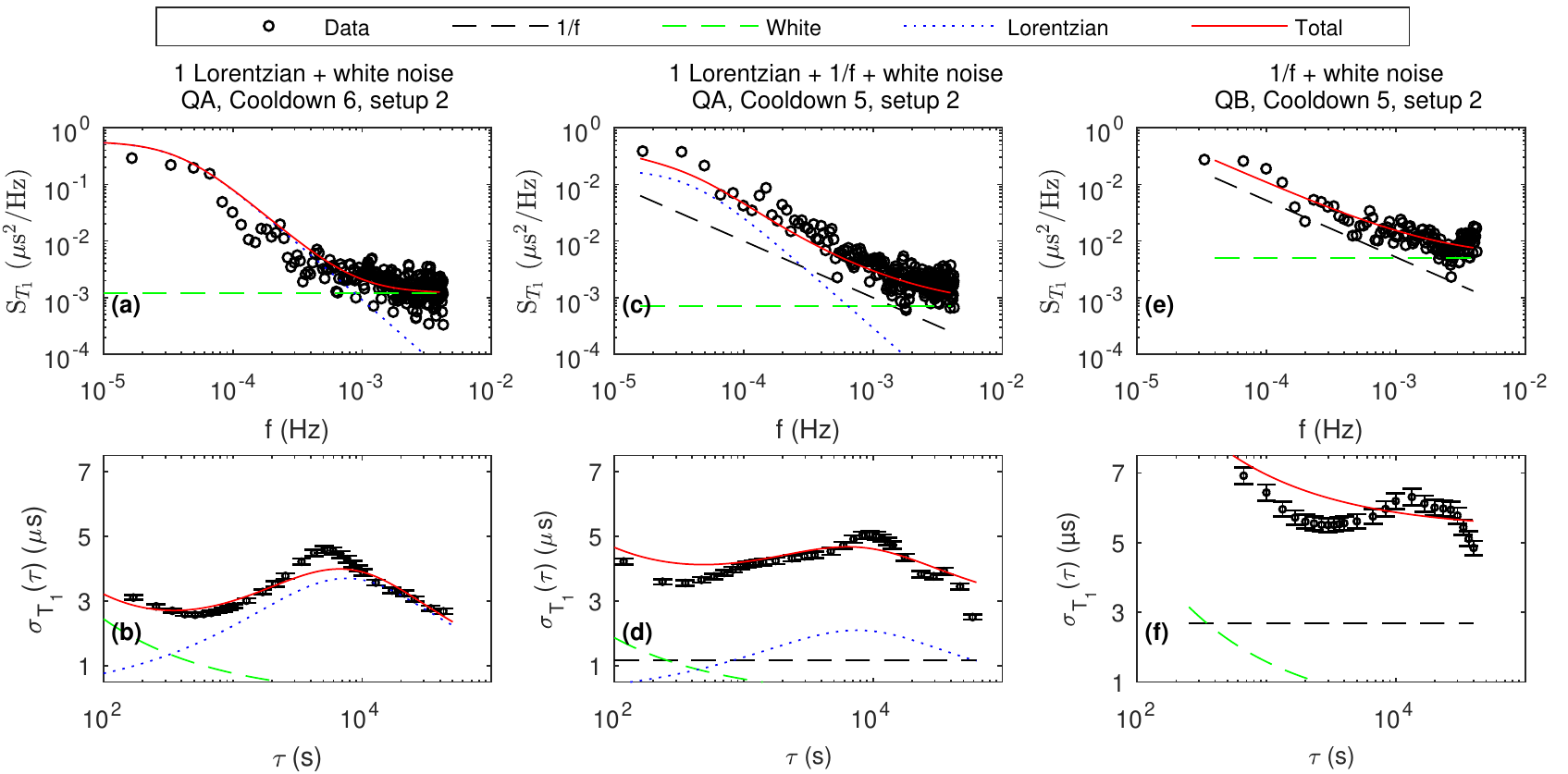}
    \caption{$T_1$ fluctuation fits to alternative noise models. {\bf (a)}, {\bf (c)}, and {\bf (e)} show the FFT of $T_1$ fluctuations, while {\bf (b)}, {\bf (d)}, and {\bf (f)} show the overlapping Allan deviation analysis of $T_1$ fluctuations. 
    {\bf (a)} and {\bf (b)} show a dataset with a clearly pronounced Lorentzian characteristic (QA Cooldown 6, cf.~\cref{fig:combo}i,l), fit to a noise model consisting of a single Lorentzian and white noise. 
    {\bf (c)} and {\bf (d)} show a dataset with a less pronounced Lorentzian characteristic (QA Cooldown 5, cf.~\cref{fig:combo}h,k), fit to a noise model consisting of a single Lorentzian, $1/f$, and white noise. 
    {\bf (e)} and {\bf (f)} show a dataset with a less pronounced Lorentzian characteristic (QB Cooldown 5, cf.~\cref{fig:combob}h,k), fit to a noise model consisting of $1/f$ and white noise.
    }
    \label{fig:altmodels}
\end{figure*}

In \cref{fig:altmodels}, we examine three sets of data and show alternative noise models to describe them. These alternative models consist of,
\begin{itemize}
    \item {\bf\cref{fig:altmodels}a--b:} single Lorentzian + white noise 
    \item {\bf \cref{fig:altmodels}c--d:} single Lorentzian + $1/f$ + white noise 
    \item {\bf \cref{fig:altmodels}e--f:} $1/f$ + white noise 
\end{itemize}

Beginning with a dataset that shows a strong Lorentzian characteristic (\cref{fig:altmodels}a--b). We examine whether the noise can be described just by a single Lorentzian. The model parameters are $h_0= \SI[per-mode=symbol]{1.2e-3}{\square \micro \second \per \hertz }$, $A^{Lor1}=\SI{6}{\micro \second}$ and $1/\tau_{0}^{Lor1}=\SI{250}{\micro\hertz}$. In this example, the Allan deviation (\cref{fig:altmodels}b) is very well described. However, the agreement with the FFT (\cref{fig:altmodels}a) is worse, with a consistent over-estimation of the Lorentzian noise and under-estimation of the white noise. In the main text, we favoured the two-Lorentzian model as it produces a better agreement across both the FFT and Allan deviation analysis methods.

Next, in \cref{fig:altmodels}c--d, we consider a dataset where the Lorentzian peak is less prominent. Consequently, it is possible to model the $T_1$ fluctuation as the sum of a single Lorentzian ($1/\tau_0=\SI{250}{\micro\hertz}$ and amplitude of \SI{4.3}{\micro\second}), $1/f$ (amplitude of \SI{1e-12}{\square\micro\hertz}), and white noise. In this example, the noise spectrum is well modelled, while the Allan deviation shows discrepancies at the lowest and highest times. 

Finally, in \cref{fig:altmodels}e--f, we examine a dataset with the least prominent Lorentzian characteristic. Here, it is possible to model the $T_1$ fluctuation as the sum of just $1/f$ (amplitude of \SI{5.2e-12}{\square\micro\hertz}) and white noise. In this example, again the noise spectrum is well modelled, while the Allan deviation shows clear structure that is not captured by the $1/f$ noise alone. 

The transition between multiple Lorentzians and $1/f$ has been studied before\cite{nugrohonoise}, in which work it was highlighted that whether there are sufficient Lorentzians to superimpose to a pure $1/f$  depends on the density of TLS. Here, we emphasize that the two-Lorentzian model within the main text better captures all details of the data. Therefore, we fitted to two-Lorentzians, because that model was able to describe the data across all cooldowns.

\begin{figure}
    \centering
    \includegraphics[width=0.95\columnwidth]{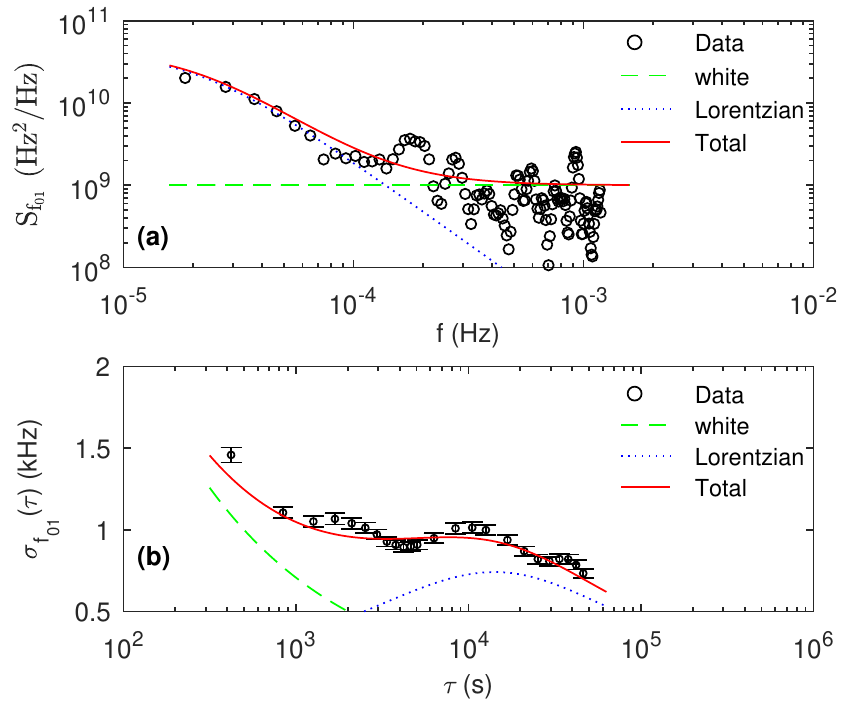}
    \caption{{\bf Frequency fluctuation data from qubit A, cooldown 3} (cf.~\cref{fig:c3}). {\bf (a)} Overlapping Allan deviation of $T_1$ fluctuations.
    {\bf (b)} Welch-method spectral density of $T_1$ fluctuations. In both plots, the $T_1$ fluctuation is fit to the sum of white noise and a single Lorentzian, where the amplitudes are the same for both types of analysis.
    }
    \label{fig:Ram1lor}
\end{figure}

The use of alternative noise models can also be extended to the qubit's frequency noise. In \cref{fig:Ram1lor} we fit the qubit's frequency fluctuations to a single Lorentzian. The model parameters are $h_0=1.0\times10^{9}$ Hz$^2$/Hz, $A^{Lor1}=\SI{1.2}{\kilo\hertz}$ and $1/\tau_{0}^{Lor1}=\SI{133}{\micro\hertz}$. 
Here, the Allan deviation (\cref{fig:Ram1lor}b) is very well described by the single Lorentzian. Equally the FFT is well described by the Lorentzian; however, in the FFT (\cref{fig:Ram1lor}a), the white noise is significantly over-estimated. In the main text, we favoured the $1/f$ model in order to enable comparison with other $1/f$ models. Physically, this was motivated by the broadband nature of the TLS dispersive coupling.

\subsection{Qubit-TLS coupling}

Within the main text, there are data sets which show revival features in time within measurements of the qubit relaxation (see \cref{fig:T1shapes}c). These features arise due to the coherent coupling between the qubit and a single TLS, described by the Hamiltonian\cite{Muellertlf},
\begin{equation}
    \frac{\hat{H}}{h}=-\frac{f_{01}}{2}\sigma_z-\frac{f_{\mathrm{TLS}}}{2}\tau_z+\frac{g_{\mathrm{TLS}}}{2}\sigma_x\tau_x,
\end{equation}
where $h$ is Planck's constant, and $\sigma_i$ and $\tau_i$ correspond to the Pauli matrices for the qubit and the TLS, respectively.
Due to this coupling, the qubit excited state can hybridize and form two, almost degenerate, states. The coupling strength $g_{\mathrm{TLS}}$  can be extracted from measuring the energy relaxation decay of the qubit and fitting it to
\begin{equation}
\begin{aligned}
    \left<\sigma_z(t)\right> = {} & \left< \sigma_z \right>_\infty +  a_{\downarrow,1}e^{-\Gamma_{\downarrow,1} t} \\ & + 
    a_{\downarrow,2}e^{-\Gamma_{\downarrow,2} t} \\ & + a_{osc}\cos(2\pi f_{osc}t)e^{-\Gamma_{osc}t}
    \end{aligned}
\end{equation}
where $\left<\sigma_z\right>$ is the expectation value of the Pauli matrix for the qubit, $\left< \sigma_z \right>_\infty$ is the zero-temperature equilibrium value, $a_{\downarrow,k}$ and $\Gamma_{\downarrow,k}$ are the amplitude and decay rate from the two excited states $k$ ($k=1,2$) to the ground state, and $a_{osc}$, $f_{osc}$, and $\Gamma_{osc}$ describe the amplitude, frequency, and decay rate of an oscillation in $\left<\sigma_z\right>$. These parameters can be rewritten in terms of coupling, $g_{\mathrm{TLS}}$, and detuning, $\delta f = f_{01} - f_{\mathrm{TLS}}$
\begin{align}
    f_{osc} &= \sqrt{g_{\mathrm{TLS}}^2+\delta f^2} \\
    a_{osc} &= \frac{g_{\mathrm{TLS}}^2}{g_{\mathrm{TLS}}^2+\delta f^2}
    \label{TLSqubiteq}
\end{align}

From this model we find a coupling rate of \SI{4.8}{\kilo\hertz} for the data in the main text. By assuming an electric dipole coupling between the qubit and the TLS, we can calculate a lower bound on the length of the electric field line, $x$, using\cite{Martinisloss}
\begin{equation}
    x = \frac{2d}{h g_{\mathrm{TLS}}}\sqrt{E_c h f_{01}} \approx \SI{39}{\micro\meter},
\end{equation}
where $d=\SI{1}{\angstrom}$ is the assumed TLS dipole length.

Additionally, from \cref{TLSqubiteq} and the data in \cref{fig:T1shapes}c, we find a TLS coherence time of approximately \SI{100}{\micro\second}. Such a lifetime is considerable larger than those found within the tunnel barrier of phase qubits\cite{Shalibotls}. However, it is strongly dependent on the coupling strength to the qubit. In absence of coupling\cite{Faorointeracting}, the phonon-limited relaxation time of a TLS is approximately \SI{1}{\milli\second}. 

The main text also shows TLS switching rates as low as \SI{75}{\micro\hertz}. These low switching rates are important in the general context of understanding TLS dynamics. Generally, measurements of charge noise\cite{Kafanovnoise} are used to determine the switching rates of TLS. Those measurements found a minimum switching rate of $\gamma_{min}$ ~=~\SI{100}{\hertz} and a maximum switching rate of $\gamma_{max}$ ~=~\SI{25}{\kilo\hertz}. Combining these leads to a TLS switching ratio ($P_{\gamma}~=~1/{\rm ln}(\gamma_{max}/\gamma_{min})$) of 0.18, a value in agreement with some experiments\cite{niepce2019high,Burnettsust}, although other studies have found lower values\cite{Graafsuppression}. A lower value of $P_{\gamma}$ can be obtained if $\gamma_{min}$ is smaller. From $T_1$ data we find switching rates ranging from \SI{75}{\micro\hertz} to \SI{2}{\milli\hertz}. Therefore, even the fastest rate is slower than those found in charge noise studies. This demonstrates that superconducting qubits are excellent probes of the TLS and highlights the need for further study of TLS dynamics. 
\\
\subsection{Local vs.~non-local origins}

\begin{figure}
    \centering
    \includegraphics[width=0.95\columnwidth]{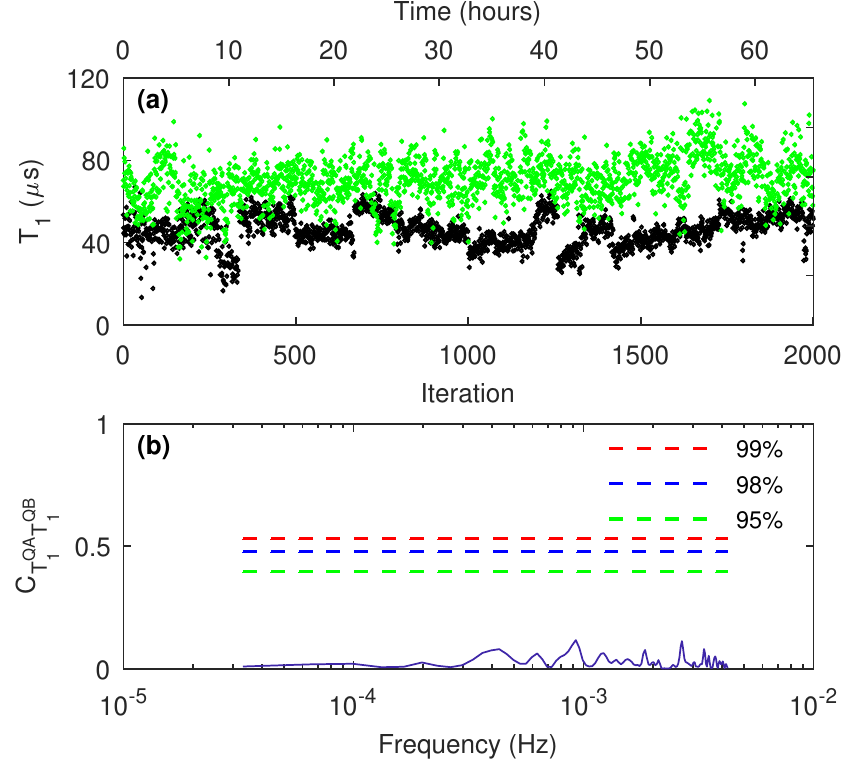}
    \caption{{\bf Magnitude-squared coherence analysis of $T_1$ fluctuations in both qubits.} {\bf (a)} Multiple $T_1$ measurements performed simultaneously on qubits A (black) and B (green). The data consists of 2000 consecutive $T_1$ measurements that lasted a total duration of \SI{2.36d5}{\second} (approximately 65 hours). {\bf (b)}~Magnitude-squared coherence of the data in  (a); the dashed lines represent the significance levels obtained from statistical bootstrapping. 
    }
    \label{fig:xcorr}
\end{figure}

Within the main text, a simultaneous measurement of both qubits is performed to examine whether the observed fluctuations in $T_1$ are local to each qubit. To assess this, we calculate the magnitude-squared-coherence of the two data sets (shown in \cref{fig:xcorr}). This examines how much the $T_1$ of qubit A corresponds to the $T_1$ of qubit B. The value of the magnitude-squared coherence is normalized to between 0 and 1, where 1 relates to completely correlated (at that frequency). In \cref{fig:xcorr} there are also dashed lines representing the threshold levels for significant correlation. These thresholds are calculated by statistical bootstrapping (repeatedly examining the magnitude-squared coherence of randomly re-sampled sets of one of the data set vs. the other original data set). The data in \cref{fig:xcorr} is clearly far below these thresholds, as would be expected for uncorrelated noise that is local to each qubit. 

\end{document}